\documentclass[conference,10pt]{IEEEtran}

\usepackage{cite}

\usepackage{graphicx}
\usepackage{ifpdf}
\ifpdf
  \usepackage{epstopdf}
\fi

\usepackage[cmex10]{amsmath}
\usepackage{amsthm}
\usepackage{amssymb}
\usepackage{cases}
\usepackage{bm}

\usepackage[caption=false,font=footnotesize]{subfig}
\usepackage{fixltx2e}

\usepackage{url}

\begin{document}
\newcommand{\tabincell}[2]{\begin{tabular}{@{}#1@{}}#2\end{tabular}}
%
\title{A Block Regression Model for Short-Term Mobile Traffic Forecasting}
%
%
%

\author{\IEEEauthorblockN{Huimin Pan, Jingchu Liu, Sheng Zhou, and Zhisheng Niu}
	\IEEEauthorblockA{Tsinghua National Laboratory for Information Science and Technology\\
		Department of Electronic Engineering, Tsinghua University\\
		Beijing 100084, China \\
		Email: \{phm13, liu-jc12\}@mails.tsinghua.edu.cn,  \{sheng.zhou, niuzhs\}@tsinghua.edu.cn\\}
}

%
%



\maketitle

\begin{abstract}\
Accurate mobile traffic forecast is important for efficient network planning and operations. However, existing traffic forecasting models have high complexity, making the forecasting process slow and costly. In this paper, we analyze some characteristics of mobile traffic such as periodicity, spatial similarity and short term relativity. Based on these characteristics, we propose a \emph{Block Regression} ({BR}) model for mobile traffic forecasting. This model employs seasonal differentiation so as to take into account of the temporally repetitive nature of mobile traffic. One of the key features of our {BR} model lies in its low complexity since it constructs a single model for all base stations. We evaluate the accuracy of {BR} model based on real traffic data and compare it with the existing models. Results show that our {BR} model offers equal accuracy to the existing models but has much less complexity.
\end{abstract}


%
\IEEEpeerreviewmaketitle

\section{Introduction}
In recent years, the proliferation of mobile devices such as smart phones and tablets have driven the traffic growth in wireless communication networks into a fast lane \cite{vni}. The surge in mobile traffic results in more radio spectrum usage and huge energy consumption, which calls for efficient network planning and operations. However, mobile traffic is highly non-uniform in both space and time, making it difficult to perform such designs. For this reason, short-term traffic forecasting, which can provide accurate and timely information for dynamic network operations such as base station (BS) sleeping \cite{TANGO,energysaving}, is of increasing importance to current and future wireless networks.

The problem of traffic forecasting has been studied extensively in the existing literature. The characteristics and forecasting models of mobile traffic are analyzed in spatial domain in \cite{tutschku1998spatial} and \cite{lee2014spatial1}. Also, a long-term traffic forecast algorithm is proposed in \cite{longterm} to forecast the traffic variation in several forthcoming years. These macroscopic models could guide long-term network planing, but cannot provide fine-grained information for dynamic network operations.

Because of this, short-term traffic forecasting should also be investigated. A local Linear Regression ({LR}) model is used in \cite{linearregression} to forecast short-term variation of vehicular traffic. Similar techniques could be applied to networks traffic forecasting, but the specialty of mobile traffic should be considered. In this respect, exogenous network information such as user profile (e.g. age and gender) and environment information \cite{usermessage} as well as mobility signaling messages \cite{mobility} can be used. Yet these information is hard to obtain in real time during network operation. In contrast, traffic forecasting based on historical traffic data is more viable. Auto-regressive Integrated Moving Average (ARIMA) model is a well-known tool in time-series analysis. A Seasonal {ARIMA} ({SA}) model is proposed in \cite{SAmodel} to forecast mobile traffic with high accuracy. Although the SA model provides satisfactory accuracy, it suffers from slow training and forecasting. To provide faster forecasting, Support Vector Machine (SVM) is used in \cite{online} and combined with fuzzy model in \cite{fuzzy} to construct traffic forecasting model. Nevertheless, both ARIMA and SVM-based models still have high complexity because they train model and forecast traffic independently for each BS.

In this paper, we analyze the characteristics of real mobile traffic and propose a \emph{Block Regression} ({BR}) model to forecast traffic in wireless communication networks. This model reflects the daily repetitive variation of mobile traffic, and offers equal accuracy with existing models such as {SA} model. At the same time, this model is formulated using linear regression and can be trained quickly with convex optimization algorithms. Above all, only a single model is constructed for all BSs in the network, greatly reducing the model complexity. This is the key difference between this model and many other existing models. That is also why we named our model with the word ``block''.

The rest of this paper is organized as follows. In Section \ref{dataset}, we introduce the traffic dataset and analyze its characteristics. In Section \ref{model}, we describe the proposed {BR} model and introduce the traffic forecasting algorithm based on this model. The proposed model is evaluated and compared with the existing model in Section \ref{result}. And the paper is concluded in Section \ref{conclusion}.

\section{Dataset Description and Traffic Characteristics}\label{dataset}
\subsection{Dataset Description}
Our dataset is provided by a Chinese company and contains the hourly data traffic of about 1000 GSM BSs from a major city.  Both uplink and downlink traffic is recorded and the unit is in Terabyte. The time span of the traffic log is one month.

\subsection{Data Cleaning}
Due to faults in the log system, there are missing values or negative values in the original dataset. Before our analysis, we clean the dataset by neglecting the traffic data of BSs which have fault data points. This leave us $N=697$ BSs for analysis. For each BS, we extract its uplink traffic data in a consecutive 14-day period and store it as follows:

 \begin{equation}\label{slidingwindowfunction}
 \bm{t}^{\{i\}} = [t_{1}^{\{i\}} , t_{2}^{\{i\}} , ... , t_{L}^{\{i\}}] ,
 \end{equation} 
where $t_{j}^{\{i\}}$ represents the uplink data traffic of $i$th BS at $j$th hour. There are $336$ hours in $14$ days, so the length of traffic vector is $L=336$. We store the vectors in matrix form, we can get a $N\times L$ matrix $\bm{T}$ as follows:

\begin{equation}
\begin{aligned}
 & [P^{\{tw1\}}_{1} , P^{\{tw1\}}_{2} , ... , P^{\{tw1\}}_{x} , A^{\{tw1\}}_{1} , A^{\{tw1\}}_{2} , ... , A^{\{tw1\}}_{y} , B^{\{tw1\}}, C^{\{tw1\}} , P^{\{tw2\}}_{1} , \\
 &  P^{\{tw2\}}_{2} , ... , P^{\{tw2\}}_{x} , A^{\{tw2\}}_{1} , A^{\{tw2\}}_{2} , ... , A^{\{tw2\}}_{y} , B^{\{tw2\}}, C^{\{tw2\}} , ...]
\end{aligned}
\end{equation}

\begin{equation}\label{datamatrix}
\begin{aligned}
\bm{T} = \left( \begin{array}{cccccccccc}
P_1^{\{twk\}} & P_2^{\{twk\}} & \cdots & P_x^{\{1\}} & A_1^{\{1\}} & A_2^{\{1\}} & \cdots & A_y^{\{1\}} & B^{\{1\}} & C^{\{twk\}}\\
P_1^{\{2\}} & P_2^{\{2\}} & \cdots & P_x^{\{2\}} & A_1^{\{2\}} & A_2^{\{2\}} & \cdots & A_y^{\{2\}} & B^{\{2\}} & C^{\{2\}}\\
\vdots & \vdots & \vdots & \vdots & \vdots & \vdots & \vdots & \vdots & \vdots & \vdots\\
P_1^{\{N\}} & P_2^{\{N\}} & \cdots & P_x^{\{N\}} & A_1^{\{N\}} & A_2^{\{N\}} & \cdots & A_y^{\{N\}} & B^{\{N\}} & C^{\{N\}}\\
\end{array}
\right)
\end{aligned}
\end{equation}

Each row of matrix $\bm{T}$ represents the traffic volume of a single BS in 14 days. Our analysis belong are all based on this matrix.  

\subsection{Traffic Characteristics and Implications}
\begin{figure}
	\centering
	\includegraphics[width=3.2in]{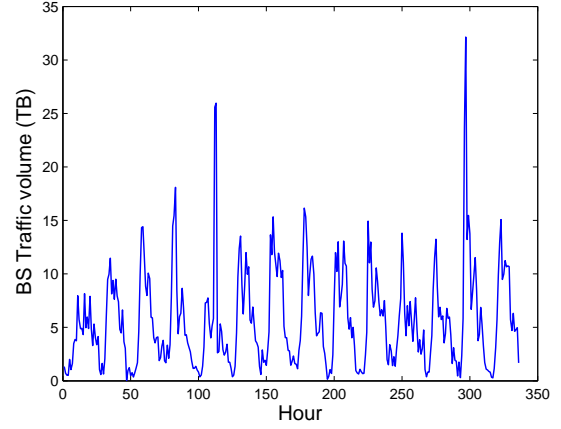}
	\caption{Hourly traffic variation of a single BS in $14$ days.}
	\label{trafficwaveform}
\end{figure}

Fig. \ref{trafficwaveform} shows the traffic variation of a single BS in a 14-day period. From the figure, we can observe a daily repetitive pattern: the traffic always goes down at night and raises at daytime. This is due to the daily dynamics of human activity. This phenomenon is common in all BSs and should be exploited when building forecasting models. In addition, we can also see that the intensity of traffic varies randomly day by day. Thus the mobile traffic is not periodic with deterministic daily pattern. For this reason, forecasting models should keep track of the daily random fluctuation in order to provide accurate forecasting.

\section{Proposed Model}\label{model}
In this section, we present our {BR} model and explain the traffic forecasting process. The whole process consists of $5$ steps: differentiation, window sliding, normalization, model training, and traffic forecasting.

\subsection{Differentiation}
In order to reflect the repetitive nature of mobile traffic, we first differentiate the original traffic with certain seasonality. The $M$-step differentiation on the original traffic at time $l$ will give us the relative deviation from the traffic value at time $l-M$. For example, if we want to take into account the daily repetition, we can set $M=24$; and if we want to include the weekly pattern, we can set $M=24\times7=168$.

Since we will get a $1\times (L-M)$ vector after $M$-step differentiation on a $1\times L$ vector. The original $N \times L$ traffic matrix $\bm{T}$ will be transformed into the $N \times (L-M)$ differentiated traffic matrix $(1-\operatorname{D}^{M})\bm{T}$.

\subsection{Window Sliding}
\begin{figure}
	\centering
	\includegraphics[width=3.2in]{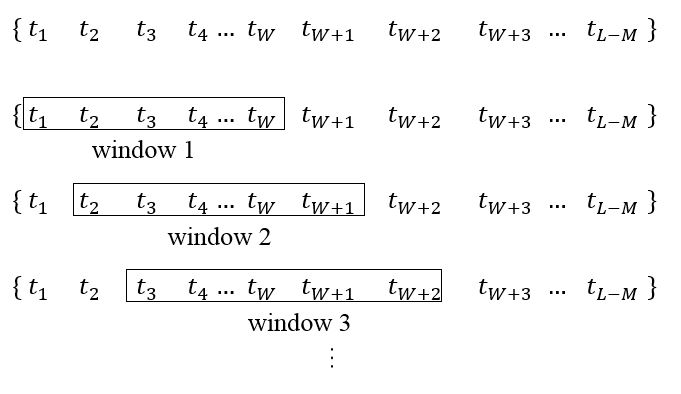}
	\caption{Illustration of window sliding on the traffic vector of one BS.}
	\label{slidingwindow}
\end{figure}

After differentiation, we employ the sliding window method to extract historic information and formulate the feature vectors for the {BR} model. Assume the window length is $W$. For each BS, we extract input and output model features as illustrated in Fig. \ref{slidingwindow}. We first align the left edge of the sliding window to the $(M+1)$-th hour in the differentiated traffic matrix, and then slide the window right-ward one hour at a time until its edge reaches the $L$-th hour. At each slide, we take the differentiated traffic values inside the window as an $1\times W$ input feature vector and the traffic value just to the right edge as the output feature. For all BSs, we will get $N_\text{s} = (L - M - W)N$ input vectors and output values. These features are then stored as rows of the input feature matrix $\bm{X}$ and output feature matrix $\bm{Y}$, respectively.

\subsection{Normalization}
In our training algorithm, each column of the training matrix $\bm{X}$ represents a set of values in each dimension of the feature space. Since the mobile traffic is highly non-uniform, $\bm{X}$ may have distinct magnitude in different feature dimensions. This may raise convergence issues to some training algorithms. Consequently, we normalize each column of $\bm{X}$ (and similarly $\bm{Y}$) to make sure it have equal variance in  different feature dimension.

The normalized dataset is represented by $\hat{\bm{X}}$ and $\hat{\bm{Y}}$. Their entries are calculated as follows:
\begin{equation}
	\hat{X}_{ij} = \frac{X_{ij} - \mu_j^{X}}{\sigma_j^{X}}
\end{equation}

\begin{equation}
\hat{Y}_{i} = \frac{Y_{i} - \mu^{Y}}{\sigma^{Y}}
\end{equation}
where $\mu_j^{X} = \frac{\sum_{i=1}^{N_\text{s}} X_{ij}}{N_\text{s}}$ and $\sigma_j^{X} = \sqrt{\frac{1}{N_\text{s}-1}\sum_{i=1}^{N_\text{s}}(X_{ij}-\mu_j^{X})^2}$ are the mean value and the standard deviation of the $j$-th column of $\bm{X}$ respectively. Similarly, the mean value and standard deviation of $\bm{Y}$ are $\mu^{Y}$ and $\sigma^{Y}$.

\subsection{Model Training}
For simplicity, we assume the differentiated traffic data satisfies the following linear relationship: 
 \begin{equation}
 \hat{Y}_i = \theta_0 + \sum_{j=1}^{W} \theta_j \hat{X}_{ij} + \epsilon_i
 \end{equation}
where $\{\theta_j: j=0,1,\cdots,W\}$ are the parameters we need to train. Among them,$\ \theta_0$ is the intercept,$\ {\theta_1,\theta_2,...,\theta_W}$ are feature weights, and $\epsilon_i$ stands for the residual noise with expected value $0$.

The model training process finds a set of parameters $\{\theta_j: j=0,1,\cdots,W\}$ which minimizes the square error between the forecast and real traffic: 

\begin{equation}\label{cost function}
\begin{aligned}
J(\theta) = \frac{1}{2N_\text{s}} \sum_{i=1}^{N_\text{s}} (\hat{Y}_i - \theta_0 - \sum_{j=1}^{W} \theta_j \hat{X}_{ij})^2 
\end{aligned}
\end{equation}

Various algorithms can be used to minimize the cost function. Among them, conjugate gradient method is widely used because of its universality and simplicity. For the same reason, we use the conjugate gradient method for cost minimization in our experiment. 

\subsection{Traffic forecasting}
We apply the trained model parameters $\{\theta_j : j=0,1,\cdots,W\}$ to forecast traffic volume. 
Specifically, if we want to forecast the traffic of a arbitrary BS $i$ at a arbitrary time $l$, we first get the $1 \times W$ normalized differentiated traffic $\{\hat{X}_{ij}: j = l-W,l-W-1,\cdots,l-1\}$ vector that correspond to BS $i$ at time $l$, and then forecast the traffic as follows:

 \begin{equation}\label{forecast_eqn}
 \begin{aligned}
  & \tilde{t}_{l}^{\{i\}} = \mu^{Y}+(\theta_0 + \sum_{j=l-W}^{l-1} \theta_j \hat{X}_{ij})\sigma^{Y} + t_{l-M}^{\{i\}}.  \\
  & (l = M+1,M+2,\cdots,L)
 \end{aligned}
 \end{equation}

Here $\tilde{t}_{l}^{\{i\}}$ represent the forecast traffic of BS $i$ at time $l$.

\section{Model Evaluation}\label{result}
In this section, we evaluate the proposed {BR} model and compare the evaluation results with the SA and LR models.

\subsection{Methodology}
We randomly select $200$ BSs and extract their traffic in a consecutive 14-day period. The traffic data is split into a training set and a test set. Traffic data in the first $10$ days is used as training set. The training data is differentiated and sampled by sliding window according to the model parameters before being used to estimate the model parameters. The accuracy of the trained model is then tested by the traffic in the remaining $4$ days. The traffic data in the test set is also differentiated and sampled by sliding window to get the test matrix.

The test matrix is then used to derive the forecast traffic as in (\ref{forecast_eqn}). The forecast accuracy of each model is evaluated in terms of the Normalized Root Mean Square Error (NRMSE). For the $i$th BS, we forecast $K$ hours' traffic and compare the forecast results with real traffic volume as follows:

\begin{equation}
\text{NRMSE}^{\{i\}} = \frac{\sqrt{\frac{1}{K}\sum_{k=1}^{K}\left(t_k^{\{i\}} - \tilde{t}_k^{\{i\}}\right)^2}}{\overline{t^{\{i\}}}}
\end{equation}
where $\overline{t^{\{i\}}}$ is the average traffic volume of BS {$i$}. Since we evaluate $200$ BSs, we will get $200$ {NRMSE} values in total. The distribution and statistics of these NRMSE values will be used as the basis for the model comparison.

\subsection{Forecasting Accuracy}
The parameters used in our evaluation are as follows. For the {BR} and {SA} models, the differential order is set to $M = 24$ to capture the repetitive daily traffic pattern. The {LR} model does not involve traffic differentiation so the raw traffic data is used. The window size for the {BR} model is set to $W_{\text{BR}}=3$, while for the {LR} model the window size is set to $W_{\text{LR}} = 72$ in order to include the historic traffic information in the last 3 days. The auto-regression and moving-average orders for the SA model are $2$ and $1$, respectively, and the seasonality is $24$.

\begin{figure}[h]
	\centering
	\subfloat[$0 < \text{NRMSE} < 1$]{
				\includegraphics[width=2.8in]{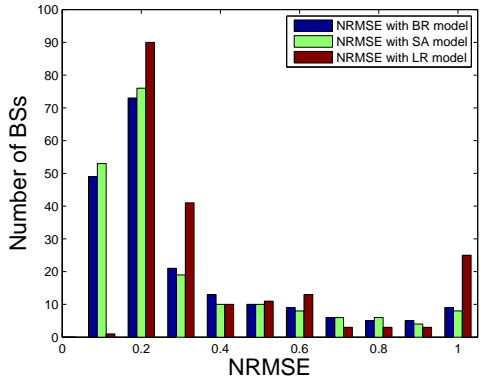} 
}
	\\
	\subfloat[$\text{NRMSE} > 1$]{
		\includegraphics[width=2.8in]{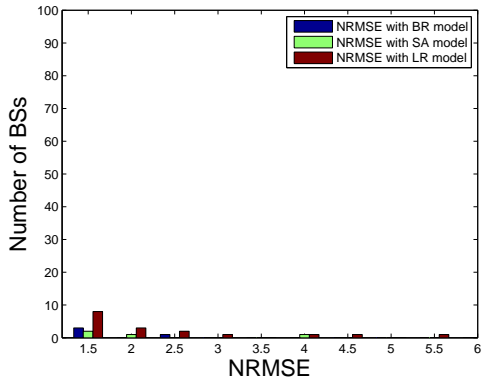} 
		}
	\caption{NRMSE histograms of different forecasting models (200 BSs).}
	\label{hist_compare}
\end{figure}

The histogram of the {NRMSE} of the $200$ BSs resulting from the {BR} model, {SA} model and {LR} model are shown in Fig. \ref{hist_compare}. It shows that {NRMSE} distribution of {BR} and {SA} models are very similar, and both of them perform much better than the {LR} model. Average {NRMSE} across all BSs for {BR}, {SA}, and {LR} models in Table. \ref{Tab_NRMSE} also shows this.

\begin{table}[h]
	\centering
	\caption{Average NRMSE for BR, SA and LR models.}
	\begin{tabular}{|c|c|}
		\hline
		Model & Average NRMSE \\
		\hline
		BR & 33.45\% \\
		\hline
		SA & 33.16\% \\
		\hline
		LR & 56.73\% \\
		\hline
	\end{tabular}
	\label{Tab_NRMSE}	
\end{table}

\begin{figure}
	\centering
	\includegraphics[width=3.2in]{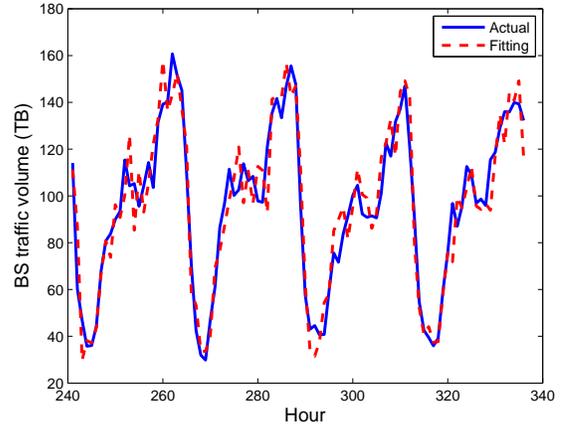}
	\caption{Traffic volume: Actual vs. Forecasts. $\text{BS}\#149, \text{NRMSE} = 0.16622$}
	\label{best_BS}
\end{figure}

\begin{figure}
	\centering
	\includegraphics[width=3.2in]{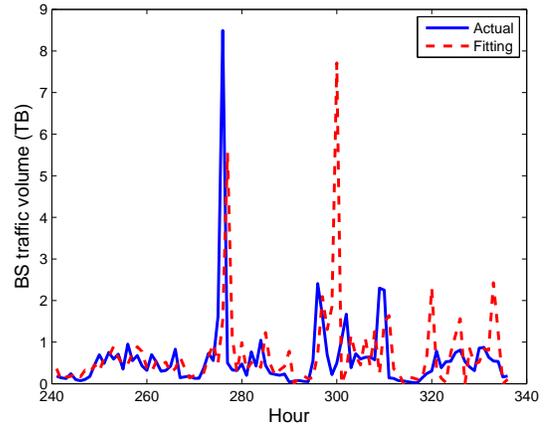}
	\caption{Traffic volume: Actual vs. Forecasts. $\text{BS}\#167, \text{NRMSE} = 1.5834$}
	\label{worst_BS}
\end{figure}

For the {BR} model, the {NRMSE} values are concentrated in the region below $0.3$, which demonstrates the accuracy of the proposed model. Note that the average {NRMSE} of BSs in the {BR} model is similar to the {SA} model; and much better than the {LR} model. 
Fig. \ref{best_BS} shows the real and forecast traffic for the BS with the lowest NRMSE. We can see that forecast traffic closely coincides with the real one.
Also, observe that there are also a few bad forecasts with {NRMSE} value larger than $1.0$. The original traffic of these bad forecasts are shown in Fig. \ref{worst_BS}.  It is obvious that the low accuracy is due to the bursty increase in the second day of forecast period, which in essence cannot be forecast based on historic traffic. 

\subsection{Seasonality}
\begin{figure}
	\centering
	\includegraphics[width=3.2in]{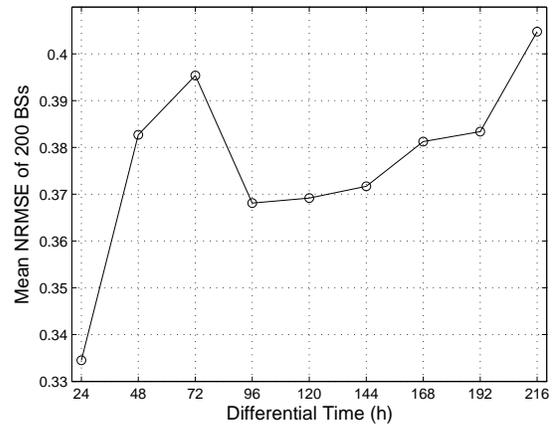}
	\caption{Average NRMSE of 200 BSs with different model seasonality.}
	\label{differential_time}
\end{figure}

The seasonality of {BR} and {SA} model should be chosen to reflect certain repetitive characteristic in the traffic. Due to the daily nature of human activity, any integer multiple of $24$ hours can be reasonably used as the seasonality. However, after experiment with our model, we find that a $24$-hour seasonality is the best choice. As can be seen in Fig. \ref{differential_time}, the average {NRMSE} of all BSs is lowest when the seasonality is set to $24$ hours. The reason for this phenomenon may be that the intensity of human activity, and thus mobile traffic, varies day-by-day. Hence the traffic value $24$ hours ago is most indicative for forecasting the traffic in the current hour.

\subsection{Model Complexity}
The three models show different model complexity. The proposed {BR} model has the lowest complexity, with $W_{BR}+1$ parameters for all the BSs. The {LR} model is more complex, with $W_{LR}$\footnote{The window size of {LR} model should be much larger than {BR} model.} parameters for all the BSs. In contrast, the {SA} model has the highest complexity. It assigns each BS with a set of parameters. For each BS, it will introduce $W_{AR}+1$ parameters in Auto-Regressive step and $W_{MA}+1$ parameters in Moving Average step. So for total $N$ BSs, it needs $(W_{AR}+W_{MA}+2)N$ parameters. For illustration, the complexity of the three models are shown in Table. \ref{Tab_complexity} with numerical examples.

Compared with {SA} model, our {BR} model has a lower complexity in two aspects. First is that differential linear model is easier than {SA} model. Our model could be considered as a special ARIMA model, which regard the MA term as 0. The reason that we could simplify the model without lossing accuracy is because we use the short term relativity characteristic of mobile traffic. After differentiation, we find that traffic volume in the next hour is only related with its past 3 hours' traffic volume. So we could only use there AR term to describe traffic fluctuation. The second aspect is that we could use a single model to describe the traffic fluctuation of all BSs. Compared with {SA} model which need to build models for each BS, it is much easier. The reason that we use a single model can have comparable accuracy with a set of {SA} models can be explained by the spatial similarity of mobile traffic. Here spatial similarity means that wireless communication traffic in different BSs and different areas follow similar variation tendency. So this is the key point why our model could work well for wireless communication traffic forecasting. As the number of BSs will be huge in real networks, low-complexity models will be more preferable than more complex models like {SA}.

Based on above two points, we can get a model with much lower complexity compared with some other existing time series forecasting models. And for those time series which have spatial similarity and short term relativity, our {BR} model will work well.

\begin{table}[h]
	\centering
	\caption{Model complexity of {BR}, {SA}, and {LR}, with examples of parameter number.}
	\begin{tabular}{|c|c|c|c|}
		\hline
		Model & Complexity & \# of Paramters & \tabincell{c}{\# of Parameters \\ used in evaluation} \\
		\hline
		BR & Low & $W_{BR}+1$ & $3+1=4$ \\
		\hline
		LR & Medium & $W_{LR}$ & $72+1=73$ \\
		\hline
		SA & High & $(W_{AR}+W_{MA}+2)N$ & \tabincell{c}{$200 \times (2+1+2)$ \\ $=1000$ }\\
		\hline
	\end{tabular}
	\label{Tab_complexity}	
\end{table}

\section{Conclusion}\label{conclusion}
In this paper, we propose a Block Regression model for traffic forecasting in  wireless communication networks. This model reflects the daily repetitive pattern of mobile traffic and has low model complexity. It could represent the traffic characteristics of all BSs with a single model without losing much accuracy. The reason that this simplify works is based on the spatial similarity characteristic of mobile traffic. 

We test this model on real mobile traffic data from a major city in China and compare the result with the local Linear Regression model and Seasonal ARIMA model. The result shows that our model has much higher accuracy to local Linear Regression model. ARIMA model is widely used in time series analysis and widely recognized to be one of the most accuracy forecasting model. Our result shows that our {BR} model have equal accuracy with {SA} model. But our model have much lower complexity compared with {SA} model, which will have significant meanings in real mobile traffic forecasting.

\section*{Acknowledgment}
This work is sponsored in part by the National Basic Research Program of China (973 Program: No. 2012CB316001), the National Science Foundation of China (NSFC) under grant No. 61201191, No. 61321061, No. 61401250, and No. 61461136004, and Hitachi Ltd.





\IEEEtriggeratref{13}

\end{document}